\documentclass[final,twocolumn,prl,aps,floats,amsfonts,amssymb,showpacs]{revtex4}
\usepackage{amsmath}
\usepackage{graphicx}

\begin{document}

\title{Convective and absolute Eckhaus instability leading to modulated waves in a finite box}

\author{Nicolas Garnier}
\altaffiliation[Present address: ]{Center for Nonlinear Science, School of Physics,
Georgia Institute of Technology, Atlanta, Georgia 30332-0430.}
\author{Arnaud Chiffaudel}
\email{arnaud.chiffaudel@cea.fr}
\author{Fran\c{c}ois Daviaud}
\affiliation{Service de Physique de l'\'Etat Condens\'e, DSM, CEA Saclay, CNRS SNC 2005, 91191 Gif-sur-Yvette, France}

\date[Published in Phys. Rev. Lett. {\bf 88} (13) 134501]{ April 2002}


\begin{abstract} 

We report experimental study of the secondary modulational instability
of a one-dimensional nonlinear traveling wave in a long bounded
channel. Two qualitatively different instability regimes involving
fronts of spatio-temporal defects are linked to the convective and
absolute nature of the instability. Both transitions appear to be
subcritical. The spatio-temporal defects control the global mode structure.

\end{abstract}

\maketitle

The Eckhaus instability \cite{Eck:65} is one of the major secondary
instability of nonlinear patterns. While the Eckhaus dynamics for steady
patterns (e.g., Rayleigh-B\'enard) is fast, dealing with local
wavelength nucleation or annihilation \cite{krazim85}, it presents slow
evolution of traveling modulated waves in the traveling-wave-pattern
case. Such modulated patterns are believed to be essential for the
description of the transition to phase and defect chaos in complex
Ginzburg-Landau models \cite{bruzim00}. Experiments on nonlinear
traveling waves are frequently carried out in annular cells
\cite{janpum92,kol92,lewpro95,liueck99,Muko:98} for the simplicity of
the underlying wave pattern, and generally consider the Eckhaus
instability close to the wave threshold. The main specificity of our
wave-system is to become Eckhaus unstable for increasing value of the
control parameter, i.e., as a first step on the route to spatio-temporal
chaos \cite{Muko:98}. In this Letter, we focus on new results in a long
rectangular cell: an homogeneous traveling-wave undergoing the Eckhaus
instability generates modulated waves. Our results reveal the rich
effect of a finite group velocity within a closed cell: we describe
quantitatively the convective and absolute modulated wave-patterns and
the associated transitions.

\paragraph*{Setup.} Our physical results concern the secondary
bifurcation of a wave pattern. We treat this system as a nonlinear wave
model: this approach does not require any connection with the underlying
physics of the convective flow. The experimental setup, its basic flow
and the nature of the primary bifurcation producing the underlying
wave-pattern have been described in detail \cite{Garnier:01}. It
consists of a thermocapillary convective flow in a long narrow channel
where an external parameter ---the horizontal temperature difference
$\Delta T$--- drives an instability toward propagating hydrothermal
waves \cite{Garnier:01,davvin93,burmuk01}. The length of the channel is
$L=180$mm. It is occasionally compared to an equivalent annular
channel~\cite{Muko:98} of perimeter $P=503$mm, i.e., a periodic boundary
condition system. The aspect ratios ensure one-dimensional patterns.

\begin{figure}[!b] 
\begin{center}
\includegraphics[clip,height=2cm]{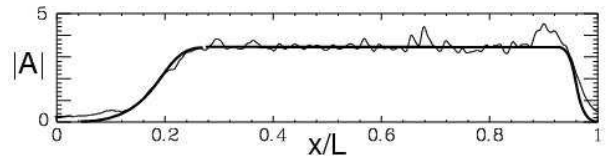} 
\end{center} 

\caption{The measured (thin line) and schematic (thick line) amplitude
profile of the single-wave system for $\Delta T = 4.75$K. The pattern is
presented as the envelope amplitude $A(x)$ of a right-traveling wave,
although both directions are equivalent. The amplitude $B(x)$ of the
minor (left) wave is negligible. For details see Fig.~2 in
Ref.~\cite{Garnier:01}.}

\label{fig:schema} 
\end{figure}

We have shown \cite{Garnier:01} this primary bifurcation to be well
described by the convective/absolute transition: a global mode is the
first structure observed when the control parameter $\Delta T$ is
increased above the absolute threshold $\Delta T_{\rm a} = 3.66$K. In
the periodic channel, waves appear at the convective threshold $\Delta
T_{\rm c}=3.1$K which can be used to build a dimensionless control
parameter $\epsilon=\Delta T / \Delta T_{\rm c} - 1$. Above complex
competition regimes between right- and left-traveling waves (e.g.
blinking states), we have shown a single wave train to become almost
uniform in the cell for $\Delta T \agt 4.5$K (Fig.~\ref{fig:schema}).
This state constitutes the basic state for the present study: we will
now focus on the secondary instability of this single wave train.

When $\Delta T$ is increased far enough from the primary onset, a
modulational instability occurs. As the group velocity is finite, the
modulational perturbations are advected. The present paper focuses on
the distinction between the convective and the absolute modulational
instability regimes and the relevance of a new object: a
front of dislocations or modulations. In periodic conditions, this
modulational instability occurs at the lowest possible wavenumber
$K_{\rm mod}=2\pi/P$~\cite{Muko:98}: it is strictly an
Eckhaus~\cite{Eck:65} instability. In the linear channel, for
simplicity, we will also refer to Eckhaus instability, although the
wavenumber of the modulational instability modes are somehow larger:
typically $K_{\rm mod} \sim 4 \! \cdot \! (2\pi/L)$.

\paragraph{Absolute instability.}

\begin{figure}[!t]
\begin{center}
\includegraphics[clip,width=8cm]{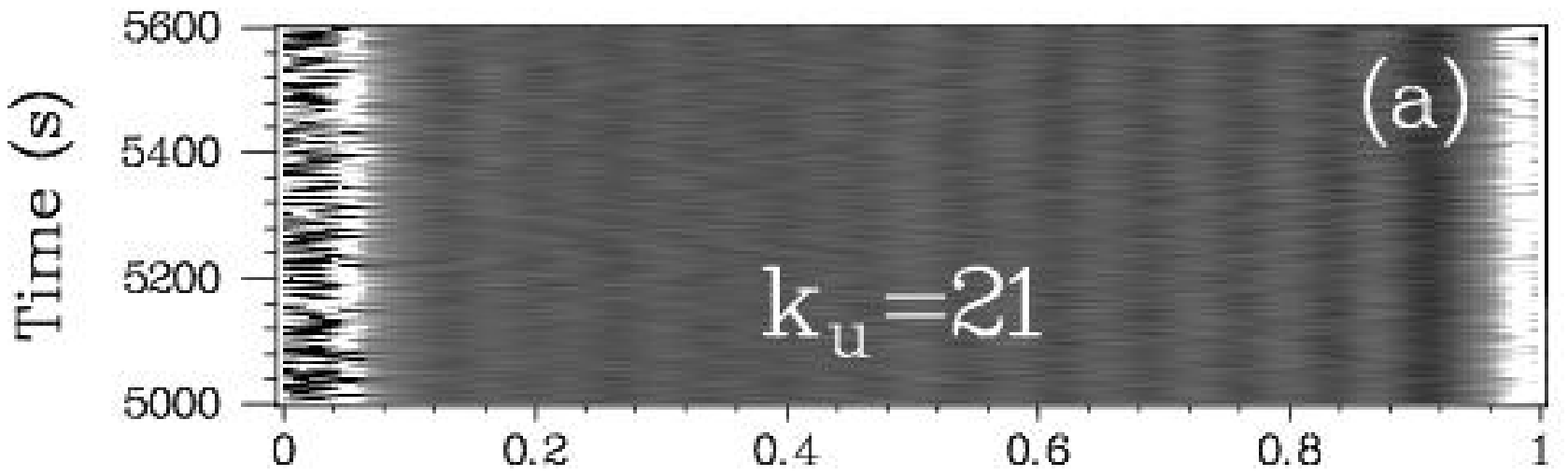}
\includegraphics[clip,width=8cm]{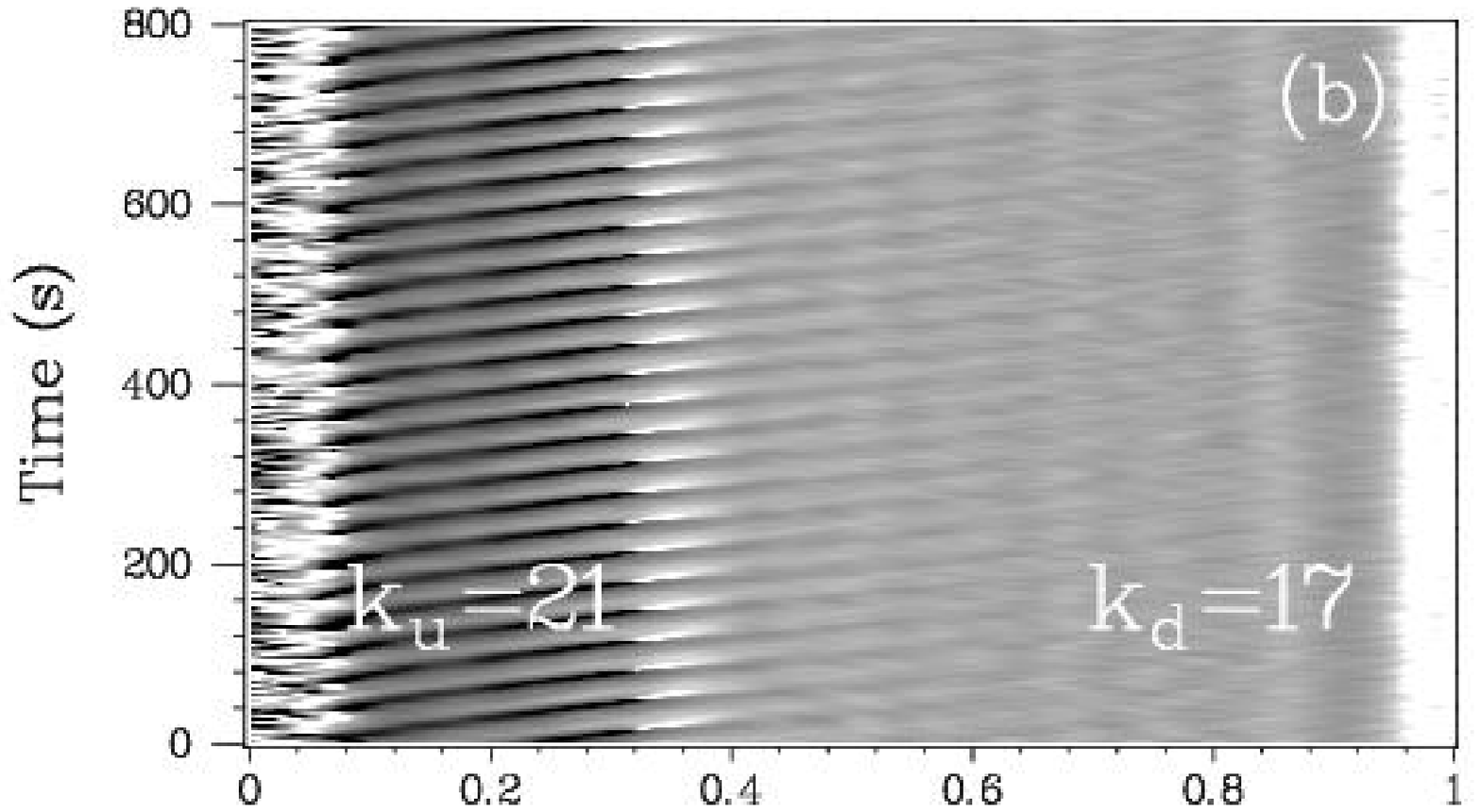}
\includegraphics[clip,width=8cm]{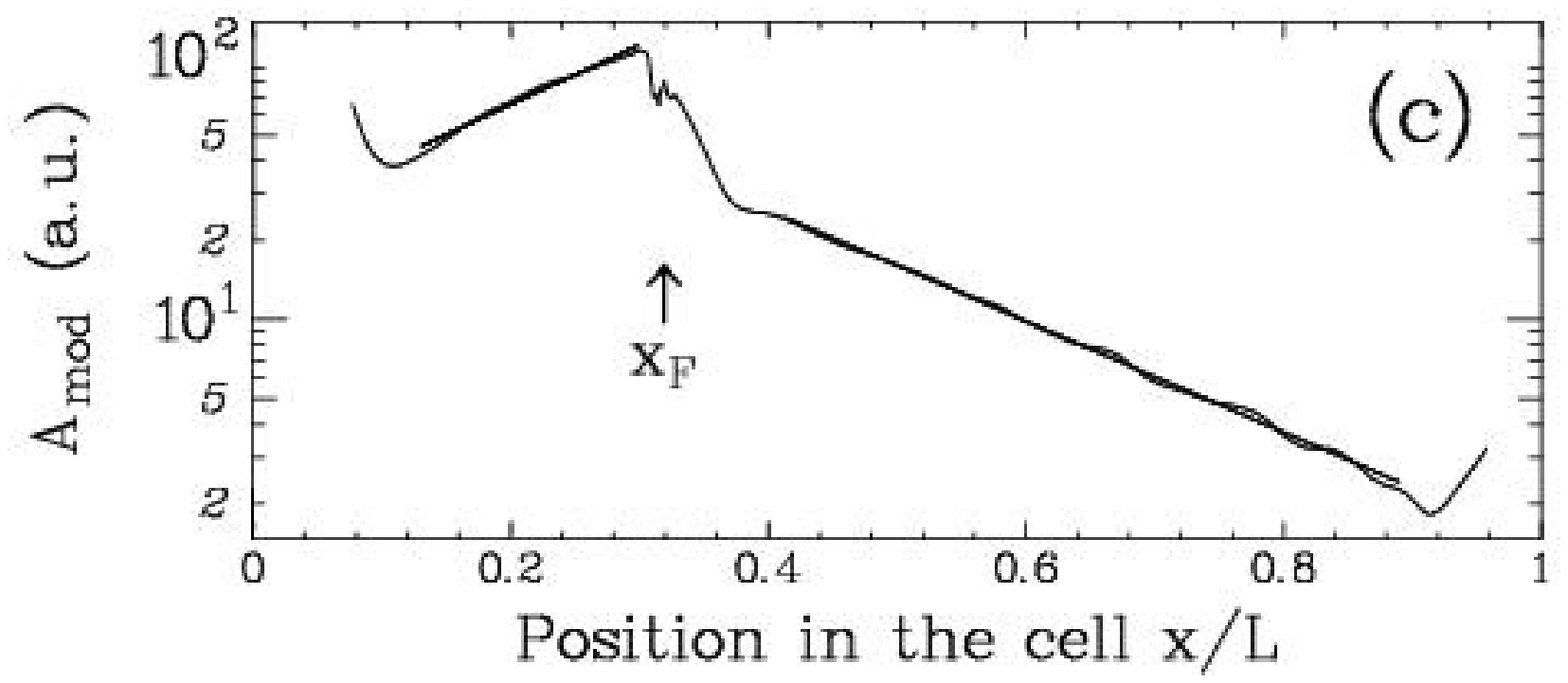} 
\end{center}

\caption{Spatio-temporal diagrams of the local and instantaneous
wavenumber $k(x,t)$ of the wave: temporally stabilized regimes for
(a) $\Delta T = 5.54$K and (b) $5.65$K. The waves propagate from left to
right. The mean wavenumber can be visually estimated by the mean gray
level and is labeled (units $2\pi/L$) in the upstream ($k_{\rm u}$) and
downstream ($k_{\rm d}$) regions. A uniform wavenumber (a) represents a
non-bifurcated state, and illustrates both stable and convective regimes
below $\Delta T_{\rm m,a}$. The modulated state (b) is the global mode
of the Eckhaus instability. Each black to white transition of the
wavenumber value at $x_F/L=0.32$ is due to a phase jump in the core of a
defect. The defect front is stable with time.
(c) By Hilbert demodulation of phase-gradient image (b) we get the
spatial profile of the amplitude $A_{\rm mod}$ of the modulation.}

\label{fig:conv-abs} 
\end{figure}

\begin{figure}[!t]
\begin{center}
\includegraphics[clip,width=8cm]{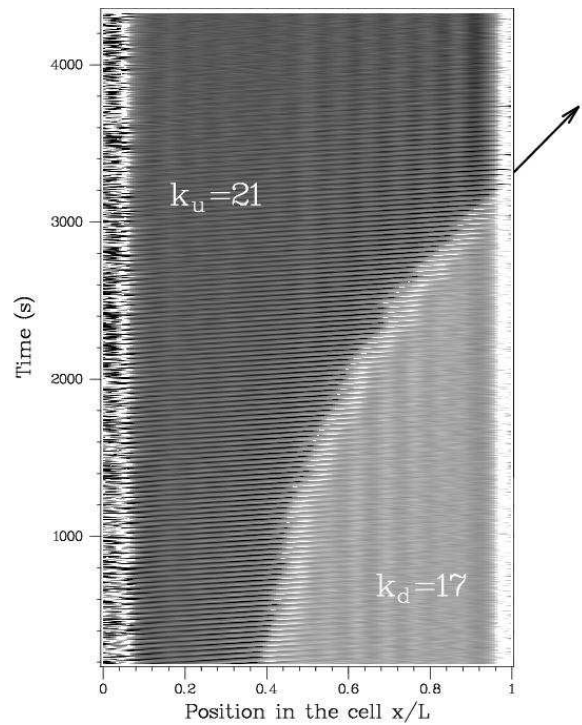}
\end{center}

\caption{Transient leading to the state of Fig.~\ref{fig:conv-abs}a.
The state has been prepared at $t=0$: a dislocation front is slowly
advected out of the cell. The modulations grow along $x$ but vanish
along $t$: this is the signature of a convective instability regime.
The arrow indicates the asymptotic front velocity.}

\label{fig:conv} 
\end{figure}

Figs~\ref{fig:conv-abs} and \ref{fig:conv} present the three states
which support our discussion. For $\Delta T > \Delta T_{\rm m,a} = (5.56
\pm 0.03)$K, the observed pattern (Fig.~\ref{fig:conv-abs}b) can be
described as a wave composed of two wave trains of mean wavenumbers
$k_{\rm u}$ and $k_{\rm d}$. The wavenumber, frequency and amplitude of
both wave trains are modulated in space and time. The wavenumber $K_{\rm
mod}$ of the modulation is of order of $|k_{\rm u}-k_{\rm d}|$. Waves
are emitted from one end of the cell with wavenumber $k_{\rm u} \sim 21
\! \cdot \! (2\pi/L)$ and propagate along the cell at the phase velocity. The
phase modulation of this wave train, traveling at the group velocity, is
spatially growing. On Fig.~\ref{fig:conv-abs}c, we clearly see the exponential
growth of the local-wavenumber modulation amplitude $A_{\rm mod}$ along
$x$. At a fixed, finite distance $x_F$ from the source-boundary, the
wavenumber modulation is so large that it allows the wavenumber to
change from $k_{\rm u}$ to $k_{\rm d}$ by time-periodic phase slips. For
$x > x_F$, the mean wavenumber is $k_{\rm d} \sim 17 \! \cdot \! (2\pi/L)$.
In this second region, the modulation is damped (Fig.~\ref{fig:conv-abs}b,c): we
conclude that $k_{\rm u}$ (resp. $k_{\rm d}$) waves are unstable (resp.
stable) with respect to modulations.

\begin{figure}
\begin{center} \includegraphics[clip,height=3.0cm]{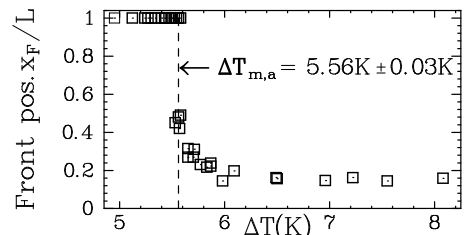} \end{center}

\caption{Spatial position $x_F$ of the dislocation front for absolutely
unstable states {\it vs} $\Delta T$. Stable and convectively unstable
states without permanent dislocation front are represented by a symbol
at $x_F=L$.}

\label{fig:front}
\end{figure}

\begin{figure}
\begin{center} 
\includegraphics[clip,scale=0.3]{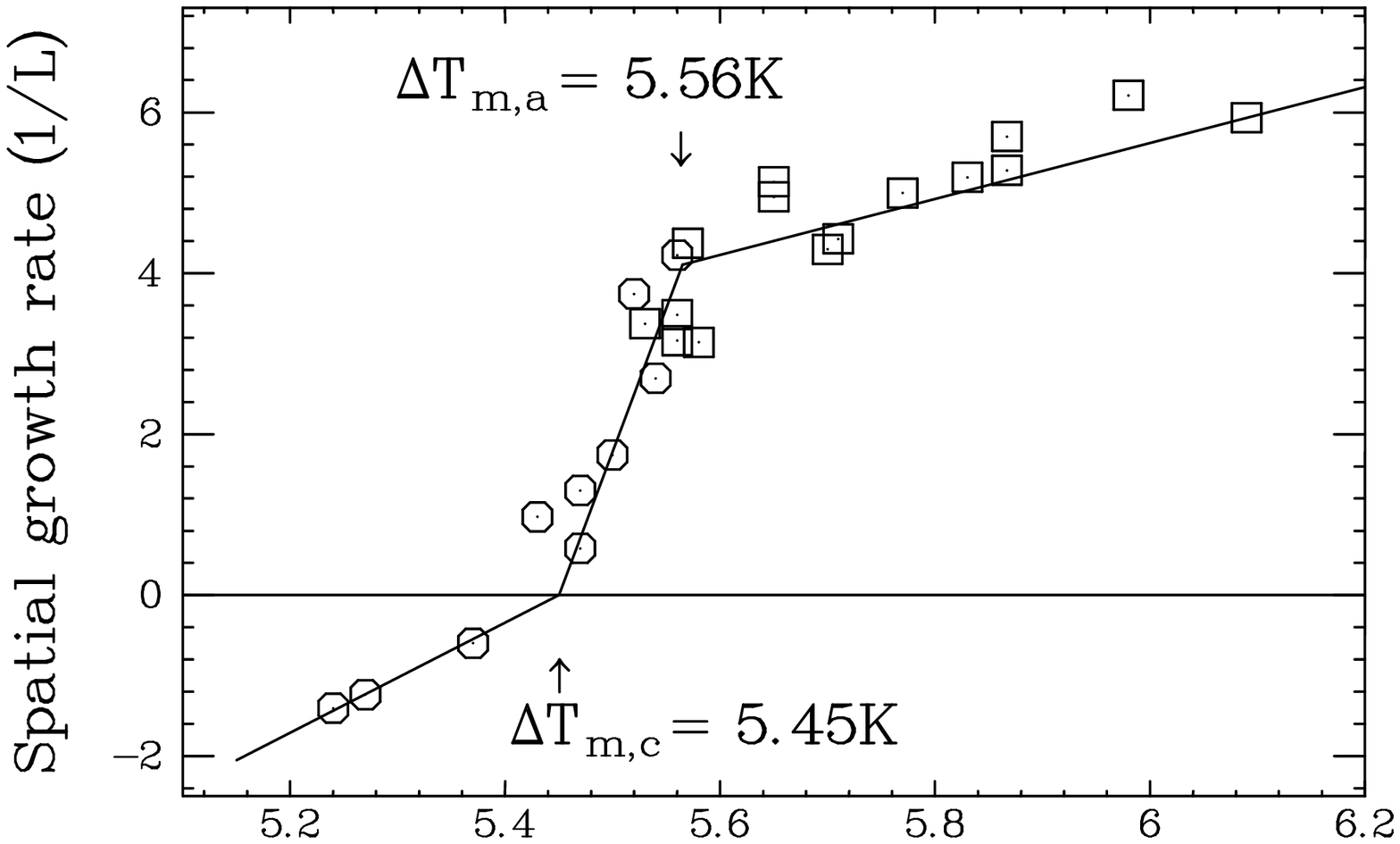} 
\includegraphics[clip,scale=0.3]{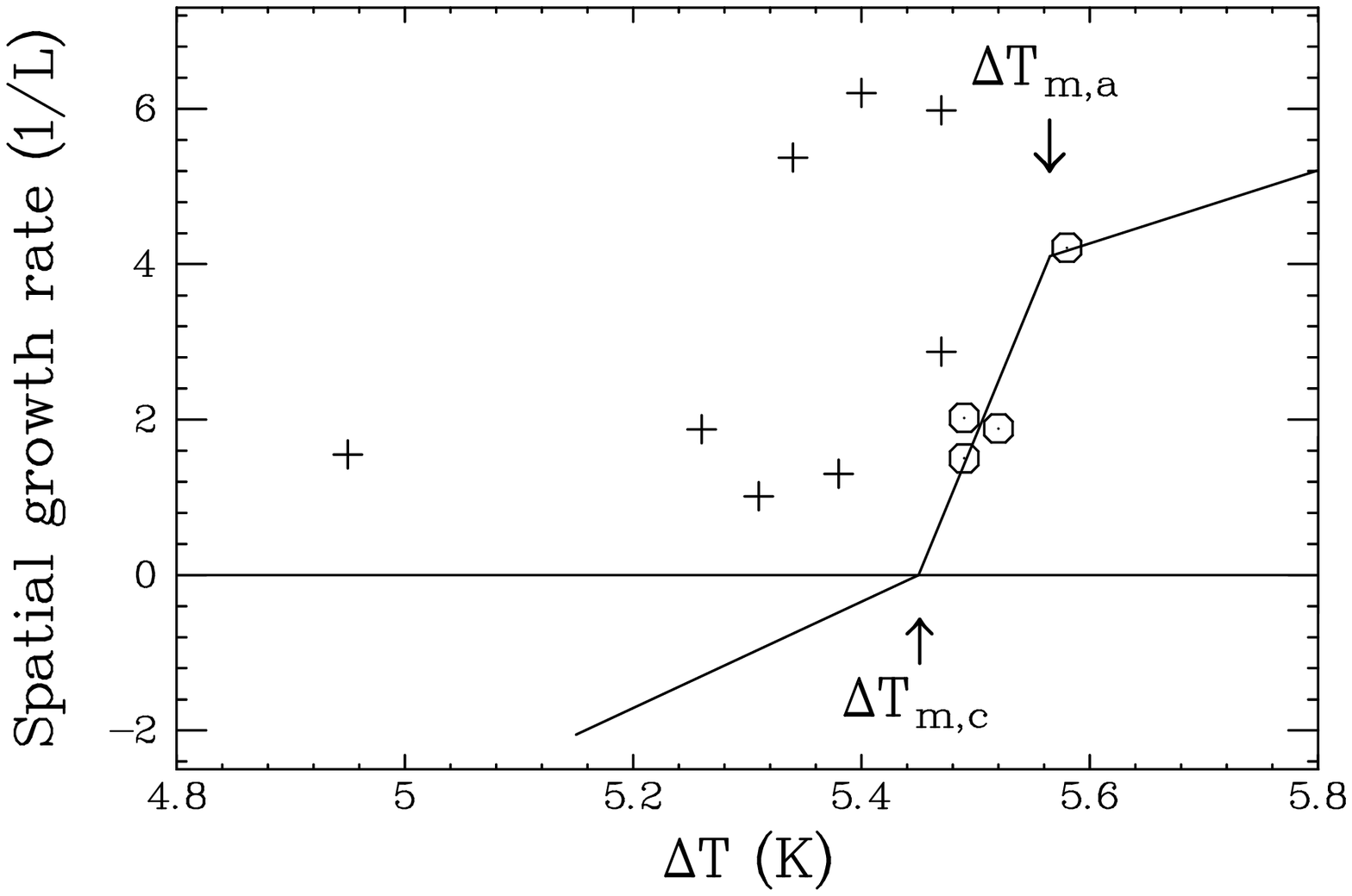} 
\end{center}

\caption{(a) Evolution of the spatial growth rate of the modulation with
the control parameter for transient ({\large $\circ$}) or steady (squares)
spontaneously modulated wave patterns. Linear fits of
the three regimes ---stable, convective and absolute--- are presented.
They intersect at $\Delta T_{\rm m,c}$ and $\Delta T_{\rm m,a}$. These
data concern the modulations of the upstream region of the cell whose
mean wavenumber is $k_{\rm u} \sim 21 \! \cdot \! (2\pi/L)$.
Corresponding data for the downstream region are negative while $\Delta
T \alt 8$K. (b)~Idem for perturbation initiated wave-packets in the
stable ({\footnotesize $+$}) and convectively unstable ({\large
$\circ$}) regimes. The solid lines reproduce the fits of (a) to allow
quantitative comparisons: the same growth rate is selected in both cases
for the convective regime.}

\label{fig:ksi}
\end{figure}

We call {\em dislocation front} the set of spatio-temporal loci where
spatio-temporal dislocations occur. For $\Delta T > \Delta T_{\rm m,a}$,
the position $x_F$ of this object is stationary; Fig.~\ref{fig:front}
shows the relation between the control parameter and the front position
which remains located in the first half of the cell whatever 
the value of $\Delta T$.
Steady dislocation fronts have been observed for traveling waves in a
Taylor-Dean experiment \cite{bot}. In general, hysteresis has not been
investigated \cite{protocol}. From the modulation amplitude profiles
$A_{\rm mod}(x)$ (Fig.~\ref{fig:conv-abs}c), we also extract the spatial
growth rate of the modulations: it is finite and positive (squares on
Fig.~\ref{fig:ksi}a).

We claim that those stationary states to be the global modes for the
modulational (Eckhaus) instability. The structure of these global modes
is very peculiar: nothing seems to saturate the modulations except the
breakup of the underlying wave-pattern, i.e., the abrupt change of the
mean-wavenumber downstream the dislocations. Similar patterns have been
numerically observed in semi-infinite \cite{Couairon:99} and closed
cells \cite{tobpro98}. Like Couairon and Chomaz \cite{Couairon:99} we
observe the nonlinear global threshold and the absolute instability
threshold to be identical. 

\paragraph*{Convective instability.}

For $\Delta T < \Delta T_{\rm m,a}$, dislocation fronts are not observed
on asymptotic states. The asymptotic regime (Fig.~\ref{fig:conv-abs}a)
is an homogeneous wave of uniform wavenumber $k_{\rm u}
\sim 21 \! \cdot \! (2\pi/L)$. However, transients obtained after
control parameter changes show traveling dislocation fronts slowly
advected out of the channel (Fig.~\ref{fig:conv}): those states are
convectively unstable states with respect to the modulational (Eckhaus)
instability. They are observed in the small gap between $\Delta T_{\rm
m,c} = 5.45$K and $\Delta T_{\rm m,a} = 5.56$K. 

For $\Delta T < \Delta T_{\rm m,c}$, asymptotic states are uniform and
dislocation fronts do not exist. Close to $\Delta T_{\rm m,c}$ very long
transients are often observed. These transient patterns (not shown) are
also slightly modulated; the modulations do not reach the critical
amplitude producing dislocations; the modulation amplitude profiles 
generally decrease (negative spatial growth rate) along the downstream
direction and slowly travel toward the upstream direction. So, the
uniform wave looks stable. The transients may last much longer than
the experimental running time, and those results have to be considered
with care.

Using a second Hilbert transform of phase-gradient data (as
Fig.~\ref{fig:conv-abs}c), we measured spatial and temporal growth rates
of the modulation. We present these data for the unstable upstream
wave train. The {\em temporal} growth rate for modulations in the
laboratory frame is negative below $\Delta T_{\rm m,a}$ and positive
above. It is also close to zero around the convective transition where
very long transients are observed. The spatial growth rate of the
upstream $k_{\rm u}$ wave train for all three regimes is presented on
Fig.~\ref{fig:ksi}a. It is positive for both unstable regimes but the
slope is seemingly different in the convective and absolute cases. It
is negative below $\Delta T_{\rm m,c}$.

\paragraph{Perturbed states.}

In order to test the above description, we perturbed the uniform
states either by plunging a thin needle in the convective layer or by
dropping a cold or hot droplet of fluid. The frequency content of those
perturbations differs from the above reported transients: the modulation
wave trains contain only a few wavelengths and appear to be advected
downstream at roughly the group velocity. All observed perturbations
show positive spatial growth rate and negative temporal growth rate in
the laboratory frame. The spatial growth rates are presented on
Fig.~\ref{fig:ksi}b. In the convective regime, the growth rate
appears to be selected at the same value as in spontaneous
transients. In the stable regime, however, the data are very dispersed
but remain positive.

\paragraph*{Discussion.}

Let us start our discussion by two important remarks:

{\em (i) The modulation amplitude $A_{\rm mod}$ never saturates:} All
observed $A_{\rm mod}$ profiles appear locally exponential along $x$. No
nonlinear saturation effect is thus observed. The occurrence of
dislocations (for $A_{\rm mod} \sim |k_{\rm u}-k_{\rm d}|$) is the only
limit to exponential growth. This is a strong argument for the Eckhaus
instability to behave subcritically in this closed cell. Remember it is
supercritical in the annular cell \cite{Muko:98}. This difference is due
to the mean wavenumber of the carrier-wave pattern. It will
be discussed elsewhere.

{\em (ii) Reflections} : The modulation wave system is a perfect single wave
system: the reflections of the modulations at the boundaries are irrelevant
since there is no possibility for reflected information to travel back to
the source.

The observation facts described above are coherent with the interpretation
in terms of convective and absolute instability. The striking point is
the positive spatial growth rates for perturbations in the seemingly
stable regime below $\Delta T_{\rm m,c}$. As for spontaneously modulated
patterns, we would expect those modulation wave-packets to decrease in
space exactly as the stable $k_{\rm d}$ wave trains do in the absolute
regime (Fig.~\ref{fig:conv-abs}c).

\begin{figure}[t]
\begin{center} 
\includegraphics[clip,height=2.5cm]{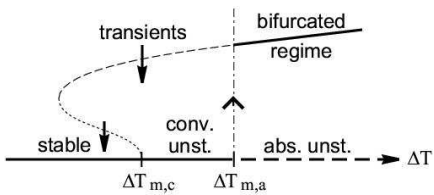}
\end{center}

\caption{Schematic representation of the observed regimes, based on the
usual representation of a subcritical bifurcation with zero group
velocity. The ordinate is only qualitative. Solid heavy lines represent
the steady states, bifurcated or not, above or below $\Delta T_{\rm
m,a}=5.56$K. The thin dashed lines may account for two different
transient modes (see text).}

\label{fig:map}
\end{figure}

Suppose that the convective instability is subcritical as suggested in
remark {\em (i)}. Then, above $\Delta T_{\rm m,c}$, the transient
evolves on an unstable branch (Fig.~\ref{fig:conv}) close to the
absolute branch (Fig.~\ref{fig:conv-abs}b). However, below $\Delta
T_{\rm m,c}$, a second unstable branch co-exists, which can be reached
only by perturbing the flow: this description can be supported by the
schematic Fig.~\ref{fig:map} inspired by zero group velocity
instabilities. These branches present very different patterns: The upper
branch exhibits extended modulations over the whole cell, with slow
evolution and, for high enough amplitudes ---the generally observed case
above $\Delta T_{\rm m,c}$---, dislocation fronts. The lower branch
exhibits fast-traveling narrow modulation wave trains and cannot be
reached spontaneously by varying $\Delta T$.
This hypothesis can explain the very different aspect of spontaneous and
induced transients in the stable regime below $\Delta T_{\rm m,c}$. It
is also known that the shape of induced nonlinear patterns below
subcritical instabilities depends of the forcing amplitude
\cite{botcha98}, so the dispersion in Fig.~\ref{fig:ksi}b may be due
to both the effect of amplitude and the presence of the two branches.

\begin{figure}[!bt]
\begin{center} \includegraphics[clip,scale=0.3]{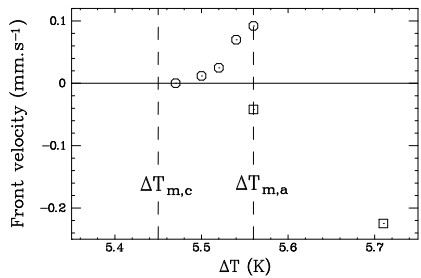} \end{center}

\caption{Front velocity around the convective/absolute transition. The
circles ({\large $\circ$}) show the velocity of dislocation fronts in
transient convective regimes below $\Delta T_{\rm m,a}$. Above $\Delta
T_{\rm m,a}$, the (negative) velocities of transient modulation fronts
invading the cell from downstream are shown by squares.
For comparison, the group velocity at wave onset is $0.90 {\rm
mm.s^{-1}}$.}

\label{fig:vfront}
\end{figure}

Another observation of the convective branch is intriguing. We record
the asymptotic velocity of the dislocation fronts between $\Delta T_{\rm
m,c}$ and $\Delta T_{\rm m,a}$, i.e., the tangent to the space-time
trajectory when the front leaves the cell (Fig.~\ref{fig:vfront}). The
observation is surprising: the closer we are to the absolute instability onset,
the faster the front moves! Then its velocity jumps below zero above $\Delta
T_{\rm m,a}$. {\it A contrario}, around $\Delta T_{\rm m,c}$, the front
velocity is zero, leading to infinitely long transients, i.e., temporal
marginality. This quantifies the experimental complexity of carrying out the
experiment around this point.
What is the meaning of the velocity jump at $\Delta T_{\rm m,a}$? Is the
convective/absolute transition also subcritical? It is probably: while
our protocol \cite{protocol} did not allow to explore all branches by
varying $\Delta T$ up and down from one state to another, a test has
been made to transit directly from an absolute state to a stable state
just below $\Delta T_{\rm m,c}$: the absolute modulation profile remains
fixed in the cell. This can be due either to hysteresis, or to the
vanishing front velocity... which makes the system marginal in this
region. This point would need to be addressed with an improved
experimental device.

Finally, in some regimes to be presented elsewhere \cite{garchi02}, the
front position $x_F(t)$ exhibits chaotic behaviors (period doubling or
quasiperiodicity): it can thus be viewed as the order parameter for the
modulational instability up to the transition to spatio-temporal chaos.

To conclude, we claim to have observed both convective and absolute
transitions for a modulational or Eckhaus instability in a long bounded
channel. The subcritical convective transition is characterized by zero
spatial growth rate and zero advection velocity for the modulated wave
pattern, which can be viewed as spatial and temporal marginality. The
absolute transition is characterized by the dynamics of dislocation
fronts. The front velocity data suggest the transition to be subcritical
as well. This question deserves a theoretical support which remains,
to our knowledge, unexplored. 


We thank A.~Casner and C.~Gasquet who participated to the data
collection and J.M.~Chomaz, J.M.~Flesselles and P.~Manneville for
fruitful discussions.


\begin{thebibliography}{20}
\bibitem{protocol} In order to get long data series close to onsets, we
often used the following protocol: the temperature gradient is first
established in the vessel, the cell is refilled and the fluid is
agitated to break the thermal gradients. Within a few seconds the
hydrothermal waves reappear without history.
\end{thebibliography}
\end{document}